\documentclass[manuscript]{acmart}
\usepackage{tabularx}
\usepackage{xspace}
\usepackage{graphicx}
\usepackage{subfigure}

\usepackage[utf8]{inputenc}
\usepackage{booktabs}
\usepackage{graphicx}
\usepackage{caption}
\usepackage{array}
\AtBeginDocument{%
  }

\setcopyright{acmlicensed}
\copyrightyear{2018}
\acmYear{2018}
\acmDOI{XXXXXXX.XXXXXXX}

\acmConference[Conference acronym 'XX]{Make sure to enter the correct
  conference title from your rights confirmation emai}{June 03--05,
  2018}{Woodstock, NY}
\acmISBN{978-1-4503-XXXX-X/18/06}

\begin{document}

\title{Investigating Youth AI Auditing}


\author{Jaemarie Solyst}
\email{jaemarie@cs.uw.edu}
\affiliation{%
  \institution{University of Washington}
  \city{Seattle}
  \state{Washington}
  \country{USA}
}

\author{Cindy Peng}
\email{cindypen@andrew.cmu.edu}
\affiliation{%
  \institution{Carnegie Mellon University}
  \city{Pittsburgh}
  \state{Pennsylvania}
  \country{USA}
}

\author{Wesley Hanwen Deng}
\email{hanwend@cs.cmu.edu}
\affiliation{%
  \institution{Carnegie Mellon University}
  \city{Pittsburgh}
  \state{Pennsylvania}
  \country{USA}
}

\author{Praneetha Pratapa}
\email{ppratapa@andrew.cmu.edu}
\affiliation{%
  \institution{Carnegie Mellon University}
  \city{Pittsburgh}
  \state{Pennsylvania}
  \country{USA}
}

\author{Amy Ogan}
\email{aeo@andrew.cmu.edu}
\affiliation{%
  \institution{Carnegie Mellon University}
  \city{Pittsburgh}
  \state{Pennsylvania}
  \country{USA}
}

\author{Jessica Hammer}
\email{hammerj@andrew.cmu.edu}
\affiliation{%
  \institution{Carnegie Mellon University}
  \city{Pittsburgh}
  \state{Pennsylvania}
  \country{USA}
}

\author{Jason Hong}
\email{jasonh@cs.cmu.edu}
\affiliation{%
  \institution{Carnegie Mellon University}
  \city{Pittsburgh}
  \state{Pennsylvania}
  \country{USA}
}

\author{Motahhare Eslami}
\email{meslami@andrew.cmu.edu}
\affiliation{%
  \institution{Carnegie Mellon University}
  \city{Pittsburgh}
  \state{Pennsylvania}
  \country{USA}
}

\renewcommand{\shortauthors}{Solyst et al.}




\begin{abstract}
Youth are active users and stakeholders of artificial intelligence (AI), yet they are often not included in responsible AI (RAI) practices. Emerging efforts in RAI largely focus on adult populations, missing an opportunity to get unique perspectives of youth. This study explores the potential of youth (teens under the age of 18) to engage meaningfully in RAI, specifically through AI auditing. In a workshop study with 17 teens, we investigated how youth can actively identify problematic behaviors in youth-relevant ubiquitous AI (text-to-image generative AI, autocompletion in search bar, image search) and the impacts of supporting AI auditing with critical AI literacy scaffolding with guided discussion about AI ethics and an auditing tool. We found that youth can contribute quality insights, shaped by their expertise (e.g., hobbies and passions), lived experiences (e.g., social identities), and age-related knowledge (e.g., understanding of fast-moving trends). We discuss how empowering youth in AI auditing can result in more responsible AI, support their learning through doing, and lead to implications for including youth in various participatory RAI processes.
\end{abstract}

\maketitle

\section{Introduction}

Teens are regular users and stakeholders of Artificial Intelligence (AI), often leading the way as early adopters of new AI-driven technologies. Like other stakeholders, they too experience the potential harms associated with AI, raising growing concerns about its negative impacts on young people. For example in 2024, a 14-year old tragically died by suicide, and his mother later found that he had detrimental interactions with an AI chatbot that supported and enhanced the teen's suicidal ideations \cite{roose2024can}. Youth themselves have expressed concerns with recent innovation in AI, such as with the new capabilities of generative AI (genAI) including an increase in misinformation and deep fakes. Eighty percent of teens in a recent poll believed that it was important for lawmakers to address new risks from AI \cite{booth2024teens}.

Responsible AI (RAI) efforts aim to reduce potential harms of AI while enhancing the potential benefits towards more ethical technology for a range of users by working towards fairness, transparency, accountability, and inclusivity (e.g., \cite{barocas2023fairness, kaur2020interpreting, diakopoulos2016accountability}). Recently, there has been an increasing focus on \emph{participatory} approaches that engage end-users and other relevant stakeholders in the design (e.g., co-design \cite{mucha2020co, donia2021co}) and ongoing evaluation (e.g., collaborative decision making \cite{harrington2019deconstructing, le2015strangers}, user-engaged AI auditing \cite{shen2021everyday, devos2022toward}) of human-centered AI \cite{delgado2023participatory}. 
However, while participatory RAI research has started to explore youths' inclusion in AI design (e.g., \cite{skinner2020children, irgens2022designing, foster2023co}), this emerging work is often limited to initial design stages. A gap remains in engaging youth in the AI lifecycle, including continuous evaluation and governance of AI. 

There is an argument for including youth in RAI processes throughout the AI lifecycle ``not in spite of [their] age, but specifically because of it,’’ one teen suggested in an interview with \textit{Time} \cite{booth2024teens}. Due to youth 1) being early adopters of AI (i.e., among the first users and ways of using AI different from adults) \cite{madden2013teens}, 2) having experiences growing up with AI-driven systems that are unique to this time of innovation (i.e., adults have had different experiences with technologies and do not have the same insight as current youth) \cite{xu2024growing, ali2021children}, and 3) expressing an interest in contributing to AI fairness (i.e., wanting to engage with design and evaluation of AI) \cite{wang2023treat, wolfe2024representation, solyst2023would, solyst2023potential}, youth are a core underexplored stakeholder in participatory RAI. Furthermore, youth have demonstrated great potential to engage in taking action toward more ethical AI. In 2020, teens in the UK protested an AI grading algorithm that disproportionately penalized lower-income students \cite{kolkman2020f}. Their public demonstrations led to increasing awareness of algorithmic inequities, ultimately resulting in the government halting the use of the AI. When supported adequately, youth can exercise agency toward more socially just AI practices. Recent research further highlights the potential of youth to engage in emerging RAI practices. For example, prior work suggests that youth are capable of recognizing and critically contemplating harmful AI bias \cite{solyst2023potential}. It has also been demonstrated that teens can conduct audits with classmates’ AI projects, identifying harmful biases and reflecting on their implications \cite{morales2024youth}.

Despite such strong evidence of youths’ potential to actively participate in RAI, there remains a gap in pathways for their inclusion in RAI as a stakeholder group. Questions remain around potential barriers youth face when actively engaging with RAI efforts regarding their technical fluency, moral reasoning, and a need for protection from potentially harmful topics that could come up \cite{solyst2023potential}. To overcome barriers to participation, prior work has shown that youth can be supported with critical AI literacy to foster more complex understandings and engage in critical discourse about AI ethics \cite{solyst2023would, wang2023treat, wolfe2024representation, ali2021children}. In this paper, we take a step to address these gaps by looking at youth engagement in user-engaged auditing, a participatory RAI approach in which AI users leverage their own knowledge and lived experiences to uncover harmful biases and other problematic AI behaviors \cite{deng2023understanding, shen2021everyday, devos2022toward}. We ask:

\begin{itemize}
  \item RQ1: How do youth go about actively finding problematic behavior in common AI technologies?
  \item RQ2: How does supporting youth with critical AI literacy scaffolding impact their AI auditing capabilities?
\end{itemize}

In a workshop study with 17 teens, we found evidence that youth could conduct insightful AI audits. Youth audited AI based on their domain expertise (e.g., hobbies and passion), lived experiences (e.g., social identities), and age (e.g., knowledge of popular trends). These audits were enhanced with critical AI literacy interventions such that youth had more developed hypotheses about what and how to audit. We also observed how youth thoughtfully wrote reports for problematic AI behavior that they discovered using a prototyped youth-inclusive AI auditing tool, as well as shared their ideas for how they think user feedback should be handled in the design of future systems. We end this work with discussion about how youth engaging with AI auditing could lead to creating AI more responsibly, yield potential educational benefits, and support their empowerment in the age of AI more broadly.

\section{Related Work}

\subsection{Recognizing and Supporting Youth’s Potential to Engage in Responsible AI}
Despite lacking efforts to include youth perspectives in RAI, young people have great potential to engage. Adults’ notions may hinder youth participation in RAI processes, including perceptions that young people lack the technical expertise, moral reasoning abilities, or maturity required to engage with complex socio-technical issues \cite{solyst2023potential}. Nevertheless, emerging evidence pushes back against these assumptions, suggesting that youth possess valuable lived experiences and critical perspectives that can make significant contributions to RAI. For example, Solyst et al. (2023) found that youth as young as 11 through 17 years old were sensitive to and articulate of bias in everyday AI, such as Google search results and genAI images, as well as thoughtful about nuances of AI ethics \cite{solyst2023potential}. This study also found that youth could engage in co-design of RAI processes to include their and their communities’ perspectives. Ultimately, without much scaffolding, children have exhibited skills and knowledge that are necessary to weigh in critical perspectives. Coupling with their capabilities to form and express nuanced opinions, youth have expressed \emph{wanting} to weigh in their perspectives in the decisions about AI systems that impact them and their communities \cite{solyst2023would, wang2023treat}. Questions remain about how youth may leverage their knowledge and perspectives to contribute to RAI, should they want to.

To further augment youths’ capabilities, critical AI literacy, i.e., understanding not only how AI functions but the ability to consider societal and ethical impacts \cite{wang2023treat, wolfe2024representation, solyst2023would}, is core to supporting youth in thoughtfully navigating everyday AI systems. For example, prior work has found that youth sometimes overtrust AI technologies, such as complex genAI, which has a growing presence in children’s lives \cite{solyst2024children}. This overtrust highlights the importance of fostering critical AI literacy—helping youth develop the skills to interrogate the sociotechnical implications of AI systems, alongside the technical capabilities. Emerging approaches to AI education, particularly for systemically marginalized learners, further emphasize empowerment and encourage the development of ``techno-social change agency,’’ such that youth are positioned to engage with and innovate toward equitable computing technologies \cite{scott2015culturally, coenraad2022s, dangol2024mediating, scott2016techno, li2023want, solyst2023would}.
When scaffolded the right way, youth can engage in critical discourse about AI, even without technical AI literacy \cite{solyst2025RAD}. However, often, critical AI literacy stays in the classroom due to a lack of infrastructure to include youth perspectives. Given youths’ potential and educational initiatives that support their critical thought about AI systems, there is great opportunity to include youth as active contributors in participatory RAI efforts. Building on prior literature, this work explores how youth can engage in participatory RAI as legitimate contributors to the evaluation of AI systems.

\subsection{User-engaged AI Auditing}
Recent years have seen growing prominence of AI audits as a method for uncovering biased, discriminatory, or otherwise harmful behaviors in algorithmic systems \cite{noble2018algorithms, asplund2020auditing, sweeney2013discrimination, prates2020assessing, buolamwini2018gender, hannak2014measuring, sandvig2014auditing, metaxa2021auditing, birhane2024ai}. At a high level, AI auditing refers to a process of repeatedly testing an algorithm with inputs and observing the corresponding outputs, in order to understand its behavior and potential external impacts \cite{metaxa2021auditing, birhane2024ai}. While most early AI audits are conducted by experts such as researchers and AI practitioners \cite{cabrera2021discovering, bird2020fairlearn, bellamy2018ai, birhane2024ai}, end users can often identify and raise awareness of harmful AI behaviors, often by leveraging their unique identities and daily interactions with AI systems that impact their lives \cite{deng2023understanding, lam2022enduser, devos2022toward}. For example, end-users often discover harmful biases in text-to-image generative AI systems that expert auditors fail to detect \cite{mack2024they, zhang2024partiality, shelby2024generative}. To this end, researchers in FAccT, HCI, and AI \cite{lam2022enduser, ojewale2024towards, kiela2021dynabench, deng2025weaudit}, along with practitioners from major technology companies \cite{chowdhury2021introducing, HuggingFace, AItest, ChatGPT_Feedback}, have begun exploring tools and processes to facilitate more user-engaged approaches to AI auditing.

However, the current line of user-engaged AI auditing research and practices often neglect an important group of AI users–the youth. In particular, AI auditing has so far been very recently investigated as an educational activity in fostering youths’ critical AI literacy. Morales-Navarro et al. (2024) examined how adolescents explored the limitations of classmates’ AI projects \cite{morales2024youth}. The study found that youth were able to identify and critically reflect on potentially harmful biases in AI systems made by fellow peers. Moreover, their engagement in the auditing process contributed to the development of their critical socio-ethical understanding of AI, highlighting the possible educational benefits of AI auditing. In this study, we aim to understand more open-endedly how youth go about exploring AI’s limitations across different types of systems, to further characterize their strengths and where support may bolster their engagement. 

\section{Methods}
\subsection{Participants}
Participants (N = 17), aged 13 to 17 (avg = 15.2), were recruited through community fliers distributed at community centers and through youth-centered STEM afterschool programs in a mid-sized city on the East Coast of the United States. Recruitment emphasized that no prior experience with AI was necessary. All youth ages 13 to 17 who expressed interest were invited to join the study. We decided to work with children no younger than 13, since this was the youngest age limit to use technologies that we focused on in the study. We had two groups of learners (A and B, participants assigned by indicated schedule availability) who took part in the research (see Table \ref{tab: demographic}). Most participants were girls, since one of the organizations we recruited through was an organization focused on gender representation in STEM. Fifteen of the 17 learners had some prior experience in computing. Participants received \$10 per hour as compensation for their time and contributions to the workshops.
\begin{table}
\footnotesize
\centering

\caption{Participants' demographics, showing mostly girls with some prior experience with computing topics .}
\label{tab:learners}
{%

\begin{tabular} {p{.09\linewidth}|p{.07\linewidth}|p{.09\linewidth}|p{.15\linewidth}|p{.03\linewidth}|p{.45\linewidth}}
\toprule
\textbf{Participant} & \textbf{Age} & \textbf{Gender} & \textbf{Race} & \textbf{Prior} & \textbf{Summarized Self-described Prior Experience} \\  \midrule

A1  & 17 & Female & White & Yes & Completed pre-college program on CompBio and Cloud Lab history \\ 
\hline
A2  & 15 & Female & White & Yes & Was in a robotics elective, previously took coding and CS classes \\
\hline
A3  & 14 & Female & Asian & Yes & Used AI tools like ChatGPT and Gemini \\
\hline
A4  & 14 & Female & White & Yes & Attended some STEM or computing-focused afterschool programs \\
\hline
A5  & 15 & Female & White & Yes & Took a college class on AI \\
\hline
A6  & 16 & Female & White & Yes & Completed CS classes \\
\hline
A7  & 16 & Female & White & Yes & Took AP CS and Machine Learning; part of a robotics team \\
\hline
A8  & 15 & Female & Asian & Yes & Studied Python 1 Java, and AP CS; active in STEM club \\
\hline
A9  & 13 & Male & Asian & Yes & Completed robotics and coding classes; used ChatGPT for ideas \\
\hline
B1  & 14 & Female & White & Yes & Created a complex dog robot; toured robotics facilities; participated in computing-focused afterschool program\\
\hline
B2  & 14 & Female & Black & Yes & Minimal experience coding with Python for robotics \\
\hline
B3  & 17 & Female & White & Yes & Familiarity with ChatGPT \\
\hline
B4  & 17 & Female & White & No & None \\
\hline
B5  & 17 & Female & White & No & None \\
\hline
B6  & 13 & Female & Black, White, Native American & Yes & Took CS and AI course at a college \\
\hline
B7  & 17 & Male & White & Yes & Enjoy working on CS and robotics projects \\
\hline
B8  & 15 & Female & Asian & No & None \\
\bottomrule
\end{tabular}%
\caption{
   An overview of the demographic of workshop participants.
}
\label{tab: demographic}
}
\end{table}

\subsection{Workshop Design}
Each participant attended a 2-hour IRB-approved research workshop session that integrated multiple structured activities to foster critical AI literacy and encourage reflection on AI and its (harmful) biases. Participants first completed a short \textbf{Pre Survey}, which asked demographic information, as well as about prior knowledge related to computing--\textit{``Have you had experience with computer science, robotics, or AI?’’} (yes/no), and to elaborate if so.

\subsubsection{\textbf{Introduction to AI}}The session began with an introduction to AI, where participants shared their initial definitions of AI in a group discussion. We then defined AI with a few common everyday examples. They then discussed their use of AI in daily life and considered whether AI could be wrong or potentially harmful. This part of the session helped everyone get on the same page about AI and begin reflecting on the nuanced impacts of AI.

\subsubsection{\textbf{Break the AI-Part 1}}Following this, we investigated how youth audit AI without much structured support in an activity where participants used individual laptops and were tasked to ``break'' three types of AI: DeepAI (text-to-image genAI), Google Images (curative AI), and Google Search Bar (recommendation AI).
We defined ``breaking'' the AI as making the AI output something that the youth thought was incorrect, they disagreed with, or could potentially have harmful effects on people.
We chose these different types of AI due to their ubiquity with both teen and adult users, as well as to be able to investigate differences in how youth audited systems that had different functions and applications. Throughout the activity, they documented their auditing on printed out auditing log worksheets that we created with space for them to write their inputs (open-ended), observations of outputs (open-ended), and if they thought that their input broke the AI (yes/no/unsure).
The participants were instructed not to use any personal information as inputs or prompts that would not be appropriate for school; however, they could note on their auditing log worksheets any inappropriate outputs. Researchers facilitated by probing participants’ hypotheses and rationale for specific audits in individual conversations throughout the activity, taking notes on their conversations. Afterward, participants collectively debriefed and discussed how they detected issues.

\subsubsection{\textbf{Critical AI Literacy Intervention}}
Next, to learn about AI ethics explicitly, participants had a short 15-minute critical AI literacy intervention about AI fairness through a structured presentation and guided discussions (see Appendix \ref{appendix_criticalAILiteracy}). Our goal was to understand how an educational intervention AI and its socio-technical harms could shift learners' perspective beyond viewing AI as technical systems toward perceiving AI as \textit{socio}-technical systems. The intervention defined the concepts of bias and harmful AI bias by probing the youth to discuss any bias they noticed in examples we showed, including: text-to-image outputs for ``a wedding,’’ image search results for ``secretary’’ (showing mostly White women) and search bar results for ``why are asian.’’ These examples were based on prior work \cite{solyst2023potential, devos2022toward}. We connected this activity to nuanced societal bias around social identities (e.g., how certain groups may be more adversely affected by biased AI outputs) and the ways in which AI-based decisions can yield unfair societal impacts. Discussion prompts, \emph{“How can biased AI harm people?”} and \emph{“Is all biased AI bad?”} encouraged participants to reason about the complexities of these technologies in small group discussions.

\subsubsection{\textbf{Break the AI Part 2}} To understand the potential impacts of AI education, participants then engaged in second activity of breaking AI, revisiting the same three AI systems with the same prompt, to break the AI. Through this Part 2, we explored how youths’ approaches or perceptions may have shifted having had the AI literacy intervention. Due to time constraints, Part 2 was about half the duration of Part 1. We then asked the participants to share again how they aimed to break the AI, and if their approach was different compared to in Part 1.

\subsubsection{\textbf{Youth Auditing Tool}}
Lastly, we had youth engage with a tool we prototyped to support auditing (see Appendix \ref{appendix_toolInterface}). Participants could compare and contrast text-to-image AI-generated images from two prompts, reflect on the outputs, and submit a scaffolded report for problematic AI behavior. The design of this scaffolding tool was inspired by Variation Theory that suggests contrasting outputs can help users identify critical features and develop a more nuanced understanding, as well as the recent interfaces designed by HCI researchers \cite{ling2012variation}. They were then given additional AI literacy probes to support their use of the tool, shown examples where two inputs were similar with an aspect changed to assess whether bias was present in the model outputs (e.g., `Boy with a bow' vs. `girl with a bow' and `American man with his car' vs. `African man with his car') (see Appendix \ref{appendix_toolOnboarding}). After using the tool to investigate AI bias, they prepared a report that contained 1) a score out of 1 to 5 evaluating the harm of the bias (from little harm to very extremely) reflected in the images, 2) their emotional reaction to outputs (could select multiple emojis depicting the six core emotions: surprise, anger, disgust, happiness, fear, sadness) 3) an explanation of why the bias was harmful, 4) who could be harmed with a drop down to select any identity characteristics that the harm may have connected to, and 5) how the outputs could be improved to remove or lessen the harm (see Appendix \ref{appendix_toolReport}). In a short \textbf{Co-Design Ideation Activity,} we asked youth to write sticky notes with ideas for \textit{how} audits should be handled and \textit{who} should be involved in the process.

\subsection{Data Collection and Analysis}
To capture data, we took notes on our observations and of workshop-wide, small group, and one-on-one conversations, aiming to transcribe youths’ relevant quotes in our notes. We did not record the whole session to protect the youths’ privacy. Youths’ artifacts (e.g., sticky notes from the co-design activity) and worksheets were photographed and digitized for later analysis. We recorded laptop screens without audio through each computer-based activity. To process the data from the structured tasks like the Break the AI and Youth Auditing Tool activities, we transcribed reports and worksheets onto spreadsheets for further analysis. We also went through screen recordings and noted youths’ inputs and took screenshots for our review, so that we could see their behavior even if it did not make it into a report or log, as well as compare their reasoning around prompts, reflection on outputs, and the actual outputs in our analyses.

For analysis, we conducted iterative inductive \cite{clarke2017thematic}, consensus-based \cite{hammer2014confusing} thematic analysis on all sources of data described previously. By this, the researchers noted themes individually and then through rounds of conversations with all authors, feedback, and iteratively going back through the data, we finalized our coding schemes for different types of audits and identified harms. For example, after multiple authors discussing common themes throughout youths’ types of inputs in the Break the AI task, the second author coded all inputs using the spreadsheets. Through various conversations, we finalized the themes, debating different specific data points. After this, the first author confirmed the coding of all inputs. We followed this process across youth auditing activities.

\subsection{Limitations}
There are a few limitations to note with our method. This work has a larger portion of youth who had more extensive backgrounds with computing topics due to the recruitment through STEM-focused out-of-school programs and opt in processes. Further, our N-size and study design is not a randomized control trial, so we cannot speak to statistically significant differences for each type of critical AI literacy support.

\section{Findings}
We first cover youths' backgrounds, approaches to and sensemaking when AI auditing, and how their perceptions of problematic AI changed as they were supported with critical AI literacy scaffolding from the intervention and tool.

\subsection{Perceptions about AI, Harm, and Engaging with Responsible AI}
\label{perceptions}
For context, we describe how the youth were using AI in their lives and their ideas about how AI could be wrong or cause harm. All youth in the study used AI in their lives, including social media apps (e.g., \emph{``Snapchat’’}), writing apps (e.g., \emph{``Grammarly’’}), and genAI (e.g., \emph{``ChatGPT’’}). When it came to the AI systems that we focused on in the study (search bar, image search, and text-to-image genAI), all youth had used search bar and image search, and almost all (14 out of 16) had used genAI, sometimes regularly. The participants specifically reported using genAI for school-related activities, such as PhotoMath, which supported step-by-step solutions for math, or ChatGPT to help make practice problems in preparation for exams.

When asked if AI could be wrong, participants highlighted issues such as genAI making up solutions or providing incorrect information. For example, A6 noted, \emph{``Generative AI…will just make something up,’’} while A2 mentioned, \emph{``The wording of something—if you word it differently–it can get it wrong.’’} B7 added that the AI \emph{``hallucinat[es] … it will confidently say wrong answers as right ones.’’} When asked if AI could have harmful effects on people, participants raised concerns about overreliance, loss of jobs, and misinformation. B5 suggested that \emph{``it can motivate people to not work as hard because it could basically do everything for you.’’} B1 related that, people could \emph{``cheat with AI’’} and this would have the impact of not adequately \emph{``giving people credit,’’} referring to intellectual property concerns around genAI outputs. A5 mentioned job displacement emphasizing that “it can affect jobs…they take away jobs.” A4 warned about misinformation’s influence, stating that it could \emph{``give people wrong information and change their perception.''} We note that in Group B, but not Group A, one learner (A2) brought up that AI could do harm by having \emph{``bias.’’} She then referred to an example of biased text-to-image genAI producing gender bias for images of \emph{``CEOs’’} that our research team had shared over six months prior with a different afterschool program that she participated in. Overall, most youth considered issues with AI to be its technical limitations and how its technical capabilities could have negative impacts on people. With this context in mind about their initial notions about AI, in the following sections we cover how they engaged with AI auditing in everyday technologies.

\subsection{Youths’ AI Auditing: Inputs and Interpretations of Outputs}
\label{inputs}

\subsubsection{\textbf{Inputs to Audit AI}}
Overall, all youth across both Break the AI Part 1 \& 2 were able to make the AI produce something that they did not agree with or thought was wrong. Our thematic analysis of youths’ auditing logs and auditing tool reports revealed six main topics that youth intentionally used and hypothesized about.

\textit{Controversial Topics} were particularly polarizing or debated topics, sometimes with misinformation surrounding them. These topics were input across all three types of AI. Because data collection took place within a week before the US presidential election, we observed that many youth especially had politically-based inputs, e.g., \emph{``congressional official tampering with votes for a presidential election’’} (B3, genAI). Controversial topics also included inputs like \emph{``best president’’} (B2, image search), \emph{``India and Pakistan’’} (A8, image search), or \emph{``climate change’’} (A3, image search). For these prompts, youth knew that there were strong varying opinions, which helped them hypothesize about and evaluate outputs.

\textit{Fact Checks} were inputs on topics that the youth could check the correctness of with right and wrong answers. These included inputs like \emph{``donald trump fighting in Vietnam’} (B7, image search) or \emph{``show me a picture of new jersey beach’’} (A7, genAI). Although used across all three AI types, checking falsifiable facts was most common for AI with visual output (image search and genAI), and when the youth knew the answers to the facts they input. For example, B7 knew that Trump \emph{``fighting in Vietnam’’} was an event that \emph{``didn’t happen.’’} In another instance, A8 mentioned, \emph{“I was looking for things I knew about, afterward looking for things like Pocahontas and how it portrayed Disney’s version instead of actual history.’’}

\textit{Generational Trends} were inputs that referenced topics that were trending with teen populations. Similar to falsifiable information, youth could determine if output was satisfactory based on their knowledge of generational trends. However, generational trends may be more debated, given their potentially ambiguous nature, changing definitions, and fast pace of trends. For example, both B2 and B6 who sat at separate tables (i.e., they did not directly converse with one another in the session) input \emph{``brain rot’’} into the genAI. B2 specifically noted that she chose to try out this input because it was \emph{``a term only people [her] age use.’’} Other participants tried out extremely recent trends from social media, such as \emph{``mango mango mango’’} (B4, image search), which was a popular TikTok reference at the time of the study. These topics were used across all three types of AI, sometimes focused on celebrities, memes, and trending phrases. Youth often hypothesized that AI would not be caught up on such fast moving trends.

\textit{Open-ended Exploration} inputs with AI with visual outputs generally lacked specificity, e.g., \textit{``bugs’’} (A1, image search), and were sometimes particularly abstract or inputs that were difficult to conceptualize (e.g., did not have a physical form or clear representation), such as \emph{``psychology’’} (B6, genAI) or \emph{``inside AI’’} (B3, image search). Contrastingly, open-ended topics were more exploratory and non-specific with search bar, given the functionality of the AI. Search bar inputs included \emph{``Can I make a’’} (A5) and \emph{``Why is my’’} (A4), and \emph{``When will’’} (A9). Across all types of AI, we observed that youth often did not necessarily have hypotheses about open-ended inputs, as they were led by their open-ended curiosity, but they could still find issues in the outputs.

\textit{Identity-related Topics} emphasized facets of identity, which often had social biases associated with them, such as inputs on race, gender, or specific occupations (see Figure 1 Image B). These included \emph{``daycare worker’’} (B6, image search), \emph{``why are Russians so’’} (A5, search bar), and \emph{``engineers’’} (A3, image search). These types of inputs were used across all types of AI. Sometimes, the youth would use prompts that were related to systemically marginalized identities that they could relate to themselves, e.g., \emph{``transgender woman’’} (A6, genAI), and assessed it with their knowledge and lived experience of the identity.

\textit{Creative Scenarios} included detailed, dreamed up inputs that the AI with visual outputs may struggle to produce adequately. Examples include \emph{``a mako shark surfing on a sea turtle in hawaii’’} (B1, genAI), \emph{``high school classroom with donkeys in it’’} (B5, image search), and \emph{``a meatball sub sitting on top of the empire state building in the rain’’} (A2, genAI). Youth hypothesized that such creative scenarios would have details missed by the AI or inaccurately generated.


\subsubsection{\textbf{Identified Problematic AI Behavior}}
\label{behaviors}
From thematically analyzing their interpretations of the problematic behavior in the auditing logs and auditing tool reports, we saw three types of problematic behavior that the youth identified. These included misinformation, harmful bias, and inappropriate content. Overall, of all the 32 auditing log entries where learners indicated \emph{Yes,} that they broke the AI, we, the research team, agreed that they had indeed, as well as with their reasoning about the identified problematic outputs.

\begin{figure}
    \centering
    \includegraphics[width=1\linewidth]{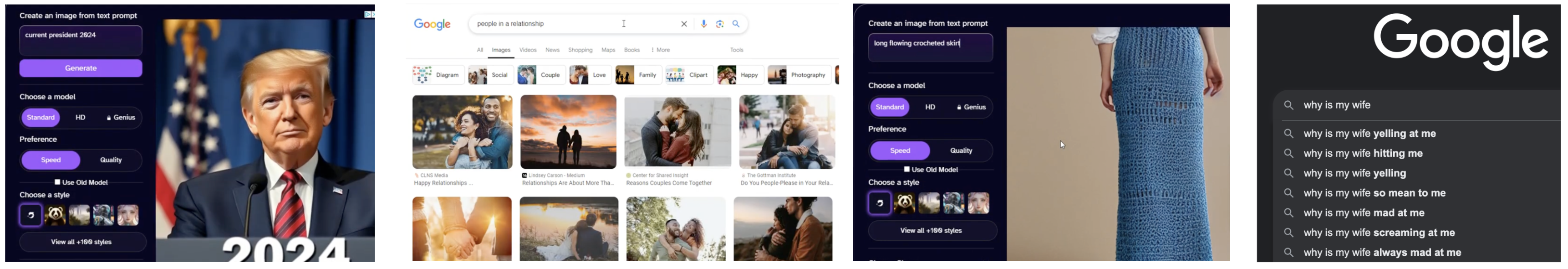}
    \caption{Examples of four types of problematic AI behavior identified from left to right representing a) misinformation shown in depiction of the president of the US in 2024, b) harmful bias in diversity of what relationships were represented in search results, c) technical limitations of crocheting output showed unrealistic stitches, and d) inappropriate content in the top few search bar autocomplete suggestions for `why is my wife.'}
    \label{fig:aisystems}
\end{figure}

\textit{Misinformation} was identified as factually incorrect or misleading outputs, in both text and visual forms. This included misleading information about elections, COVID-19, or simply incorrect information represented in images or in search bar suggestions (see Figure \ref{fig:aisystems}a). For example, with B1’s knowledge of sharks she concluded that a generated output had misinformation because the image \emph{``showed the wrong species of shark. The shark is more bulky, and the tail is short …''}. 

\textit{Harmful bias} included outputs that echoed harmful societal biases, such as stereotypes. These included problematic outputs of different inputs about systemically marginalized race, gender, or professions. For example, B5 generated an image of \emph{``Entrepreneur sitting at a desk’’} and then reflected that \emph{``the image was of a White man, which could be harmful to so many groups of people.’’}

\textit{Inappropriate Content} included outputs that were concerning or discomforting in some way. For example, A4 saw that after she entered \emph{``why is my’’} in the search bar, one of the suggested results included \emph{``why is my wife yelling at me?’’} (see Figure \ref{fig:aisystems}d). A4 remarked that the results made her \emph{``uncomfortable [because they were] very specific to issues,’’} despite her open-ended prompt.

\textit{Technical Limitations} was characterized as the outputs being generally unsatisfactory or poor quality, but not due to the other reasons above (see Figure \ref{fig:aisystems}c). For example, for some genAI outputs, participants pointed out how text in the generated images was illegible.

While there were general correlations at times between types of inputs and identified problematic AI behavior, there was also deviance. For example, while often fact checking prompts led to participants identifying misinformation in the outputs, sometimes these prompts would yield correct outputs but with harmful bias. Similarly, youth could point out different biases in the same category than they imagined. For example, the results of \emph{``transgender woman’’} did not contain the stereotypes about transgender women that A6 imagined, but rather, she noted that the output contained a \emph{``different bias, generating someone who is aryan.’’} Overall, they were not overly blinkered by their hypothesizing, and could assess AI outputs.

\subsection{Youths’ Sensemaking and Contributions in Auditing}
\subsubsection{\textbf{Using domain expertise and lived experiences}}
\label{livedexperience}
To come up with the prompts, youth often leveraged their domain expertise and sources of knowledge (e.g., life experiences, identity). These supported their critical sensemaking when hypothesizing and testing the limitations of AI, leveraging their identities and lived experiences as assets in auditing. 
This included identity characteristics, such as \emph{``Muslim’’} (A8), \emph{``African American’’} (B2), and \emph{``trans’’} (A6). Additionally, we observed them exercise their expertise. A2 had deep knowledge of crocheting, so she took a fact-checking approach in using genAI to create images of crocheted garments. On her auditing log, A2 reported that the generated image did not have \emph{``recognizable stitches''} and that it looked \emph{``glitchy,''} pointing out that the images generated were factually incorrect, given the stitches’ appearances in the generated images (Figure \ref{fig:aisystems}c). She further elaborated in a conversation that it was important and common for those who crochet to be able to tell the pattern by looking at a garment carefully, since stitches could be interpreted and replicated. Another teen, B1, had strong domain expertise in sharks. Many of her attempts to audit the AI included checking facts on information and representations of certain types of sharks. These examples show that youth can be domain experts, with knowledge far above that of an average person. We observed that these youth were particularly engrossed in the activity and enthusiastic about using their expertise to share their insights when the researchers would ask them about their approaches to auditing.

They also connected back topics from school, such as knowledge from courses on history and culture, to hypothesize and interpret outputs. A8 mentioned how she was exploring audits based on what she was learning in a class on world cultures about misperceptions: \emph{``Africa is perceived as one country, poor, and [primarily] jungle,’’} despite it being only \emph{``5\% jungle.’’} Youths’ hypotheses came from understanding of stereotypes and inequities learned from other coursework.

\subsubsection{\textbf{Transactivity and collaboration}}
\label{transactivity} We also observed the youth benefit from group sensemaking. Specifically, we saw transactivity occur, meaning collaborative learning through transaction (e.g., discussion) \cite{lewis2005transactive}, as youths’ ideas resonated with one another and reasoned about AI behavior. For example,
one participant tried out \emph{``future president''} in genAI \emph{before} the election had taken place, finding that the AI generated one specific candidate, showing a strong bias. The teen gasped in surprise and wanted to share their findings with others. This sparked curiosity in other youth in the group to try out more presidential election-related topics. The youth had unprompted discussions among each other about how surprising, interesting, and potentially harmful this AI behavior was. Transactivity occured most for youth who already had existing relationships and rapport with one another before the study, e.g., were friends or siblings who sat next to each other. Throughout the activities, we observed the youth not only expressing curiosity and surprise at problematic AI behavior, but also laughing together as they audited the AI. Through exploring the limitations of AI, this also cultivated a playful atmosphere when the youth knew each other.

\subsubsection{\textbf{Processes to handle auditing contributions to be carefully considered}}
\label{contribution}
When it came to considering how their effort towards writing reports should be handled in the co-design activity, youths’ ideas most often reflected that the original creators of AI (e.g., those based in industry) should be the ones to make AI technology better. This sentiment reflected in youths’ ideas included: \emph{``Reports should go back to whoever developed the AI to try and help them reduce bias,''} a \emph{``higher-up CEO,''} or \emph{``people researching AI.''} This suggests that youth see original creators of AI in industry as having responsibility and capability to fix the AI. Youth considered ways that the companies could prioritize surfaced issues, including processes like \emph{``put [the reports] into groups with similar issues [so that] the problems can be conquered quicker,''} or \emph{``rank reports based on how users ranked how harmful they are. Then have a way to determine how accurate that harm is.''} However, some participants expressed skepticism about companies handling things satisfactorily, one noting that they \emph{``won’t [handle the reports, because] they just want money.’’} Other participants also brought up the importance of the public opinion to continue through processing the reports, such as the \emph{``public [could] vote on how harmful/useful the reports are,''} as a way for others to continually contribute toward more democratic processes that would continue beyond just submitting a report. Overall, the youth showed both optimism and concern about how their auditing efforts would be taken into account.

\subsection{Impacts of Critical AI Literacy Scaffolding}
\label{withscaffold}


\subsubsection{\textbf{Shifting understanding of AI as socio-technical systems}}
\begin{figure}[h!]
\centering
\begin{tabular}{@{}p{0.48\textwidth}@{}p{0.48\textwidth}@{}}
\begin{minipage}[t]{\linewidth}
\centering
\renewcommand{\thetable}{\arabic{table}} 
\captionsetup{type=table,position=bottom} 
\caption{Youth inputs and identified issues showing a shift from technical toward more socio-technical audits and understandings from Break the AI Part 1 to Part 2}
\label{tab:half_width_table}
\resizebox{\textwidth}{!}{ 
\begin{tabular}{|l|c|c|c|}
\hline
\multicolumn{4}{|c|}{\textbf{Types of Prompt Inputs}} \\ \hline
\textbf{Input Types} & \textbf{Part 1 (\%)} & \textbf{Part 2 (\%)} & \textbf{Total (\%)} \\ \hline
Controversial Topics & 28.9 & 14.1 & 23.6 \\ \hline
Fact Checking & 16.7 & 4.7  & 12.4 \\ \hline
Open-ended Exploratory & 12.3 & 10.9 & 11.8 \\ \hline
Creative Scenarios & 20.2 & 14.1 & 18.0 \\ \hline
Identity-Related & 11.4 & 48.4 & 24.7 \\ \hline
Generational Trends & 10.5 & 7.8  & 9.6  \\ \hline
\multicolumn{4}{|c|}{\textbf{Identified Problematic AI Behavior in Outputs}} \\ \hline
\textbf{Problematic Output Types} & \textbf{Part 1 (\%)} & \textbf{Part 2 (\%)} & \textbf{Total (\%)} \\ \hline
Misinformation & 40.0 & 22.7 & 33.5 \\ \hline
Harmful Bias & 34.5 & 69.7 & 47.7 \\ \hline
Technical Limitations & 21.8 & 6.1 & 15.9 \\ \hline
Inappropriate Content & 3.6  & 1.5  & 2.8  \\ \hline
\end{tabular}
}
\end{minipage}
&
\begin{minipage}[t]{\linewidth}
\centering
\renewcommand{\thefigure}{\arabic{figure}} 
\captionsetup{type=figure,position=bottom} 
\includegraphics[width=\textwidth]{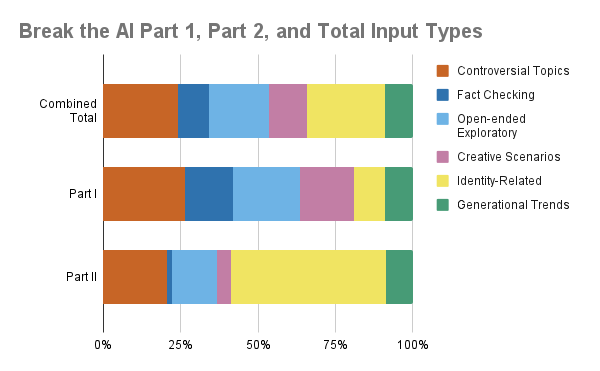} 
\caption{Graph of the input types showing a large increase of identity-related prompts from Part 1 to Part 2}
\label{fig:graph}
\end{minipage}
\end{tabular}
\end{figure}

Throughout the study, we observed an interesting shift in how youth understood and investigated AI and its potential harms, from investigating technical limitations toward more nuanced socio-technical limitations of AI. Across different types of AI literacy scaffolding, we aimed for the participants to have a better understanding of AI as socio-technical systems that could reproduce societal bias. Participants demonstrated their ability to critically assess AI behavior by providing thoughtful examples and reasoning. In particular, we saw that in Part 2, youth were more inclined to explore the limitations and problematic behavior through more intentional prompts that aimed to produce nuanced bias in AI outputs, especially those on identity-related topics. For example, in Group A, none of the teens mentioned identity-related bias as a type of harm in the initial Introduction to AI activity. However, after the Critical AI Literacy Intervention, the focus on identity prompts shifted from 0\% in Part 1 to 36\% in Part 2. Similarly, we observed an increase of identity-related prompts in Group B, from 23\% in Part 1 to 68\% in Part 2.
Although auditing specifically identity-related AI harm is not necessarily more important to audit compared to other types of harm, we see this as compelling evidence of critical AI literacy. In other words, auditing identity-related bias in AI lends insight into youths' awareness of more nuanced socio-technical AI harms (e.g., compared to checking straightforward falsifiable facts or using creative scenarios to find the technical incapabilities of text-to-image genAI). Figure \ref{fig:graph} shows the dynamics and changes of different types of input prompts across both groups in Part 1 and 2, depicting the large increase of the identity-related topics in Part 2.  

When it came to the Youth Auditing Tool, the participants’ audits became even more sophisticated with support of the onboarding materials, use of the tool’s compare-contrast feature, and scaffolded report. Notably, \emph{all} of the reports were on identity-based prompts. 
To further illustrate the impact of critical AI literacy scaffolding mechanisms, we investigate the auditing trajectories for a few of the participants, from the beginning to using the tool. B7 began the Break the AI activities in Part 1 by exploring detailed creative scenarios and political prompts such as \emph{``Evil robot AI taking over Antarctica, blowing it up to make it a large server room.’’} In part 2, he then zoomed in closer on identity-based prompts and political prompts, such as \emph{``why are African countries’’} and \emph{``the most evil democratic politician.’’} He described successfully breaking the AI by finding accurate information and stereotypes in the outputs. Using the auditing tool, he continued prompting for racial, cultural, and professional field biases in more depth. B7 rated cultural prompts as the most harmful, such as \emph{“African boat’’} vs. \emph{``American boat.’’} Interestingly, this learner also tried \emph{``realistic portrait of genius person’’} vs. \emph{``realistic portrait of dumb person.’’} He noted that while all the outputs were generally \emph{``White men,’’} the bigger issue was that he felt the AI should \emph{``not generate a dumb person,’’} i.e., that the AI should not generate outputs for insensitive inputs like this. This shows a higher level of critical thinking, even though we did not explicitly suggest this, rejecting use of AI in some scenarios altogether.

\subsubsection{\textbf{Facilitating investigation through critical comparison}}
The tool supported youth in identifying new types of harm by encouraging critical examination across different contexts through the comparison feature.  
For example, using the auditing tool, many youth highlighted the gender bias that arose when comparing professions. In one instance, A4 searched for \emph{``Soccer player’’} versus \emph{``Ballet dancer,’’} and the results depicted a majority of men playing soccer and a majority of women as ballet dancers, even though gender was not specified in the prompt. A4 remarked, \emph{``I think this could be harmful for people who participate in these activities who are not the represented gender.’’} Comparing these professions side by side instead of individually, explicitly highlighted the gender bias that the AI could reproduce.

Between the Break the AI Activity Part 1 \& 2, only two youth (B1 and B2), who were of the youngest (13 years old), of the 17 participants did not audit with prompts that would elicit social bias, despite our critical AI literacy intervention. However, with the use of the auditing tool and the opportunity to compare prompts across different contexts, they formed hypotheses and auditing based on more nuanced identity-related topics. For example, B2 explored controversial prompts and political themes such as misinformation around COVID and \emph{``Stolen elections’’} in Part 1. She explained, \emph{``I wrote this prompt because I wanted to get an output that I disagreed with [i.e., misinformation], because I do not think stolen elections exist.’’} She tested the AI's ability to handle nuanced ethical concepts. She found both outputs to be harmful since AI visually endorsed the concept. Using the auditing tool she then began exploring racial harms, as well as ones that related to facets of her own identity, such as \emph{``African American person in the neighborhood’’} vs. \emph{``White American person in the neighborhood.’’} She thought about how AI bias could harm her generation, explaining that \emph{``children could be harmed [by this output] because they grow up with the idea that specific groups [of people] have better houses than others.’’} This was also heavily influenced by the onboarding scaffolding the youth received before experimenting with the auditing tool.

\subsubsection{\textbf{Enhancing articulation through structured critique}}

We observed that the quality of the reflection and reasoning was more comprehensive with the report structure compared to the open-ended log worksheet. Comparing the youths' inputs to the output, the research team agreed that there was indeed harm \textit{present} in general in 19 out of 20 youth reports. Additionally, we saw compelling \textit{reasoning} for 18 of the 20 reports. 
Surprise and disgust were the most common emotions reported. Overall, despite a few disagreeances between the participants’ audits and the research team, we saw overwhelming evidence of their capabitilies to engage in auditing, which was enhanced through the structure of the report as scaffolding. For example, A2 found biased outputs when comparing \emph{``girls playing sports''} vs. \emph{``guys playing sports,''} which only generated images of young girls playing by themselves, whereas the depictions of the guys were \emph{``usually shirtless, with a lot of muscles playing sports with each other.''} The participant produced a fairly comprehensive audit, clearly articulating harm based on gender and physical appearance biases, and the research team agreed with the submitted report.

\section{Discussion}
In this study, we investigated the potential for youth, particularly teens under 18, to actively participate in RAI practices, specifically through AI auditing. While much of the current RAI research focuses on adult populations, this study highlights how teens, as both users and stakeholders of AI systems, can offer insights into problematic AI behaviors, despite potential notions against their inclusion in participatory RAI efforts \cite{solyst2023potential}. We find how youth go about exploring the limitations of AI and how critical AI literacy scaffolding can enhance their auditing abilities. Through prototyping and evaluating youth auditing tools, we also contribute findings with implications for designing future youth-inclusive AI auditing tools. This research suggests multiple potential gains of empowering youth to engage in auditing, regardless of prior exposure to technical education, as they draw from their expertise, lived experiences, and age-related knowledge–ultimately leading to more ethical AI and offering educational benefits for participants. Overall, this work underscores the importance of incorporating young people in the ongoing evaluation of AI systems. We elaborate more on this next.

\subsection{Youths’ Potential Contributions and Empowerment}
Youth are a critical yet underexplored demographic in user-engaged AI auditing and broader emerging participatory Responsible AI (RAI) processes. Extending prior work in user-engaged AI auditing \cite{devos2022toward, lam2022enduser, deng2025weaudit}, this study finds that youth bring perspectives from their lived experiences as young people, along with knowledge that is in line with adults’ with the right scaffolding in many cases (Section \ref{livedexperience}). Youth may also have specific domain expertise that surpasses an average person’s if they have deep knowledge in a certain topic or hobby, which we saw could be an asset in auditing \cite{ojewale2024towards}. Lastly, youth may leverage to new emerging AI more quickly or differently than adult populations \cite{madden2013teens}. This positions them as important contributors to identifying potential problematic AI behavior, especially in domains where youth-specific interests or interactions may uncover biases or unintended consequences undiscovered by adults. Importantly, although all adults have been youth previously, the technological landscape has changed drastically—growing up in an increasingly AI-driven society makes current youths' perspectives unique. Taken together, youth have critical insights that are essential to catching issues and ensuring quality interactions in RAI for younger users. One unique contribution of youth is that they have insight into how their generation is using new AI technologies, through engaging with fast-moving trends and new ways of using AI in connection to different contexts, such as school. Future work may further characterize other specific insights that youth can contribute to AI auditing.

As mentioned in Section \ref{withscaffold}, with appropriate scaffolding, youth demonstrate well-developed abilities to critically evaluate AI systems, making them valuable participants in processes such as AI auditing. Importantly, AI auditing could be a pathway for youth to exercise techno-social change agency, actively engaging in processes that shape the socio-technical systems impacting them and their communities. In addition, careful thought should be put into youth-generated feedback (e.g., reports)---the way that feedback is received and handled is crucial. As suggested by prior work \cite{birhane2024ai, deng2025weaudit}, reports should be treated as valuable contributions for supporting AI accountability and framed in ways that reinforce youth empowerment, ensuring that their voices are acknowledged and acted upon.

\subsection{Auditing as an Educational Activity}
Not only can youth contribute to auditing, but we saw that the practice of auditing can contribute to their learning. We observed that the participants engaged thoughtfully in the active task of AI auditing, and that auditing as an activity enhanced youths’ learning by doing \cite{schank2013learning}. Although specifically AI auditing may be an emerging approach to fostering critical AI literacy, there are various connections to broader educational approaches that emphasize learning by doing and students' agency through their active participation. \textit{Scientific inquiry} is one popular approach, which has especially supported students in learning STEM topics, as they experiment with topics in science to build mental models of how the natural world and engineering principles function \cite{bybee2011scientific}. Youth AI auditing may be situated as a science inquiry approach toward AI literacy. Similarly, \textit{citizen science} approaches in the classroom have empowered youth to actively participate in science and research, as they engage in collective efforts like gathering or labeling data and sharing it with scientists, learning scientific topics through practical experiences in the process (e.g., \cite{shah2016current}). There have been notable outcomes from citizen science efforts, including the public contributing to producing large datasets (e.g., \cite{raddick2009galaxy, eiben2012increased}) and influencing policy (e.g., \cite{hecker2018innovation}). Youth AI auditing can also be framed as a citizen science effort, which has both learning outcomes and potential contributions beyond the classroom. Lastly, we see connections to critical computing education efforts, such as critically conscious computing \cite{everson2022key, henrique2023creates} and culturally responsive computing \cite{scott2015culturally, scott2016techno}. These computing pedagogies emphasize fostering learners' abilities to understand, critique, and take action against systems of oppression embedded in technologies. Youth AI auditing supports all of these aspects, as they learn from engaging in auditing and can engage in enacting change through submitting audits and broader RAI discourse.


Tools that support youth in AI auditing can be designed with learning science principles. We saw in this study that using \emph{compare and contrast}, a common learning science approach applied to classrooms (e.g., \cite{rittle2009compared}, that youths’ ability to understand problematic AI outputs and audits were enhanced compared to scaffolding from a critical AI literacy intervention alone. Moreover, as we saw participants make connections to AI outputs with other school subjects, such as history class (Section \ref{livedexperience}), auditing could establish meaningful connections to school curricula by exploring how historical biases or societal narratives are encoded and perpetuated by AI systems. This aligns with broader efforts to integrate computing education across disciplines, highlighting the interdisciplinary relevance of AI literacy (e.g., \cite{barr2023cs+}). 

\subsection{Youth-Engaged Auditing Design Implications and Future Directions}
Findings from this study suggest that the future of participatory RAI should be inclusive for youth, should they want to provide their perspectives. Drawing from prior FAccT and HCI work on exploring new processes and building interfaces to support diverse people in AI auditing \cite{deng2025weaudit, lam2022enduser, cabrera2019fairvis}, future work could further explore the design and evaluation of 
youth-friendly auditing interfaces. However, many FAccT, HCI, and AI research have suggested the potential psychological harms might exhibited in conducting AI auditing and red-teaming activities \cite{lam2022enduser, shen2021everyday, devos2022toward}. Questions remain about ensuring that youths’ potential participation does not cause harm, is not exploitative, and is beneficial to the youth participants. For example, should youth be compensated for their contributions to auditing? A careful balance between protection and empowerment is needed. Auditing with educational benefits may be one way to work toward this balance. Future RAI researchers and practitioners need to pay extra attention to protect youths' mental well-being when engaging them in AI auditing, especially given that the AI output can be potentially psychologically harmful.

The design of auditing experiences and scaffolds should be carefully considered--we noted some design trade-offs from this study. First, we saw that although scaffolding with a tool supported higher quality audits and more comprehensive reflection as evidenced in the reports, the emotional experience of auditing open-endedly in the Break the AI activities versus with the tool was different. We observed that during the Break the AI activities, the experience of youth AI auditing was more playful (e.g., laughter), social (e.g., unprompted conversation), and curiosity-driven (e.g., deeply engaged and led by a broader range of interests). However, using the tool was more individual and a main reported feeling was disgust, a negative feeling. Tools that do not facilitate more open-ended social aspects of sensemaking about AI behavior in auditing may not yield as pleasant emotional experiences. Further, due to youth being an underexplored population to engage in auditing, system designs should ideally support youth with the flexibility to bring their funds of knowledge, which adults may not be able to anticipate. While we saw that scaffolding designed to support specific types of audits (e.g., identity-based in this case) yielded more extensive reflection and audit reports, such supports may unintentionally lead youth to have a narrow focus. A main strength of including youth as a population in auditing comes from discovering new AI behavior that other populations may not unearth, where audits can be youth-led or youth-specific in nature. Ultimately, these considerations underscore the importance of designing inclusive, ethical, and sustainable frameworks for youth engagement in AI evaluation.

\section{Conclusion}
In this work, we explored how teens can engage in participatory RAI, specifically AI auditing and how they can be supported with critical AI literacy scaffolding when doing so. We found that youth were able to audit using various sources of knowledge and lived experiences. Scaffolding from critical AI literacy supported their depth of audits and shifted their auditing from investigating technical limitations toward more nuanced socio-technical limitations of AI. We see opportunity for active participatory RAI practices, such as AI auditing, to be educational experiences through learning by doing, which opens up opportunities to situate RAI in learning contexts. Future processes and tools can be designed to center youth. Ultimately, this study shows evidence of how youth can be supported to engage in RAI efforts and provide their perspectives on AI fairness.

\newpage
\subsection*{Ethical Considerations}
In our IRB-approved research with youth, we took care to consider ethics, following protocols of parental consent and minor assent. At the start of the session, we introduced the idea of research and were transparent about how exactly we were collecting data. We also let participants know that they could decline any activities, discussions, and questions at any time or withdraw from the study. To align with regulations for AI use, we chose the youngest age to be 13 years old, as well as clearly instructed participants not to prompt with inappropriate or personal information. Our research approach to running a research workshop emphasized inclusive language and avoided assumptions about participants' backgrounds. We compensated youth participants at IRB-approved rates, which were deemed non-coercive by our ethics board but acknowledged the youths’ contributions to the research. We also respected the youths’ privacy by choosing not to record the whole session and aimed to respect individual comfort levels. Throughout, we prioritized transparency and participant well-being, pacing the session as needed for the youth in the workshop.

\subsection*{Adverse Impact}
There are a few potential adverse impacts, which we aim to avoid from this work. When youth engage in algorithm auditing, several potential adverse impacts must be considered. They may be exposed to harmful or inappropriate content, which can lead to emotional distress. If youth inadvertently share personal or sensitive information in auditing, this raises privacy concerns. Without proper guidance from scaffolding or an adult depending on the context, youth might develop misconceptions about how algorithms work, leading to oversimplified or incorrect conclusions. Additionally, if their contributions are not appropriately valued or compensated, exploitation could occur, diminishing the positive aspects of their participation.

\bibliographystyle{ACM-Reference-Format}
\bibliography{00-main}
\appendix

\section{Appendix}
\newpage
\subsection{Slides from Break the AI Part 1 and Part 2}
\label{appendix_BreakAI}
\begin{figure} [h!]
    \centering
    \includegraphics[width=1\linewidth]{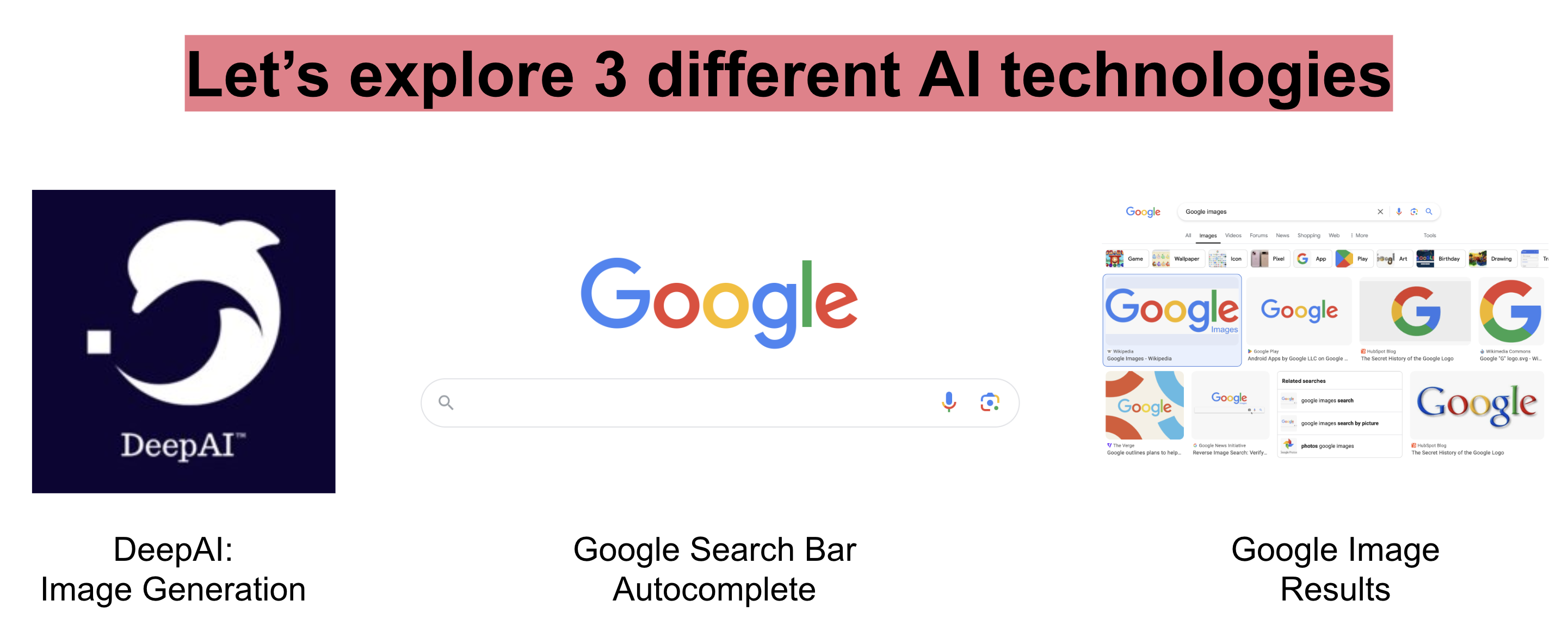}
    \caption{Students explored Deep AI, Google Search Bar, and Google Image Search to see if they can "Break the AI."}
    \label{fig:threeai}
\end{figure}

\begin{figure} [htbp]
    \centering
    \includegraphics[width=0.8\linewidth]{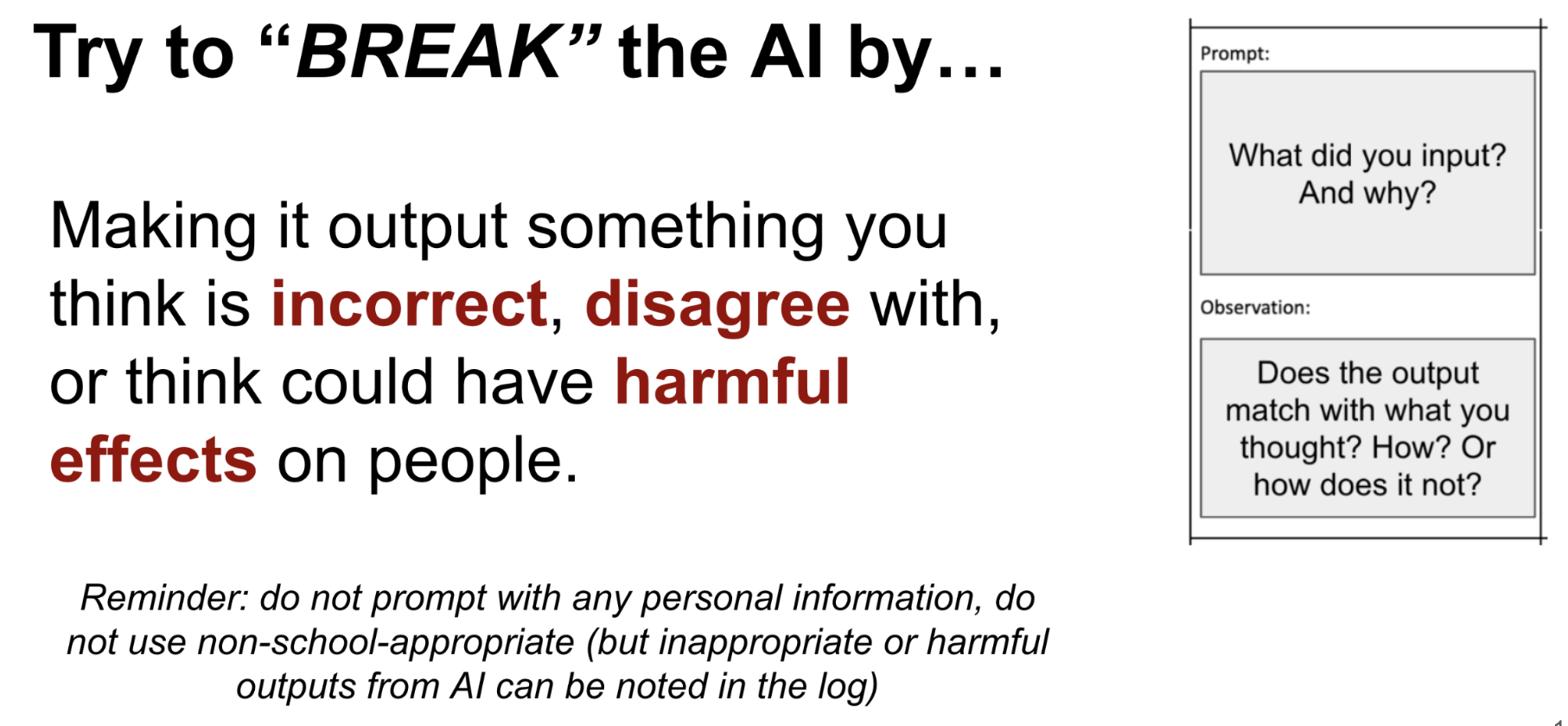}
    \caption{Break the AI method.}
    \label{fig:breakai}
\end{figure}

\begin{figure} [htbp]
    \centering
    \includegraphics[width=1\linewidth]{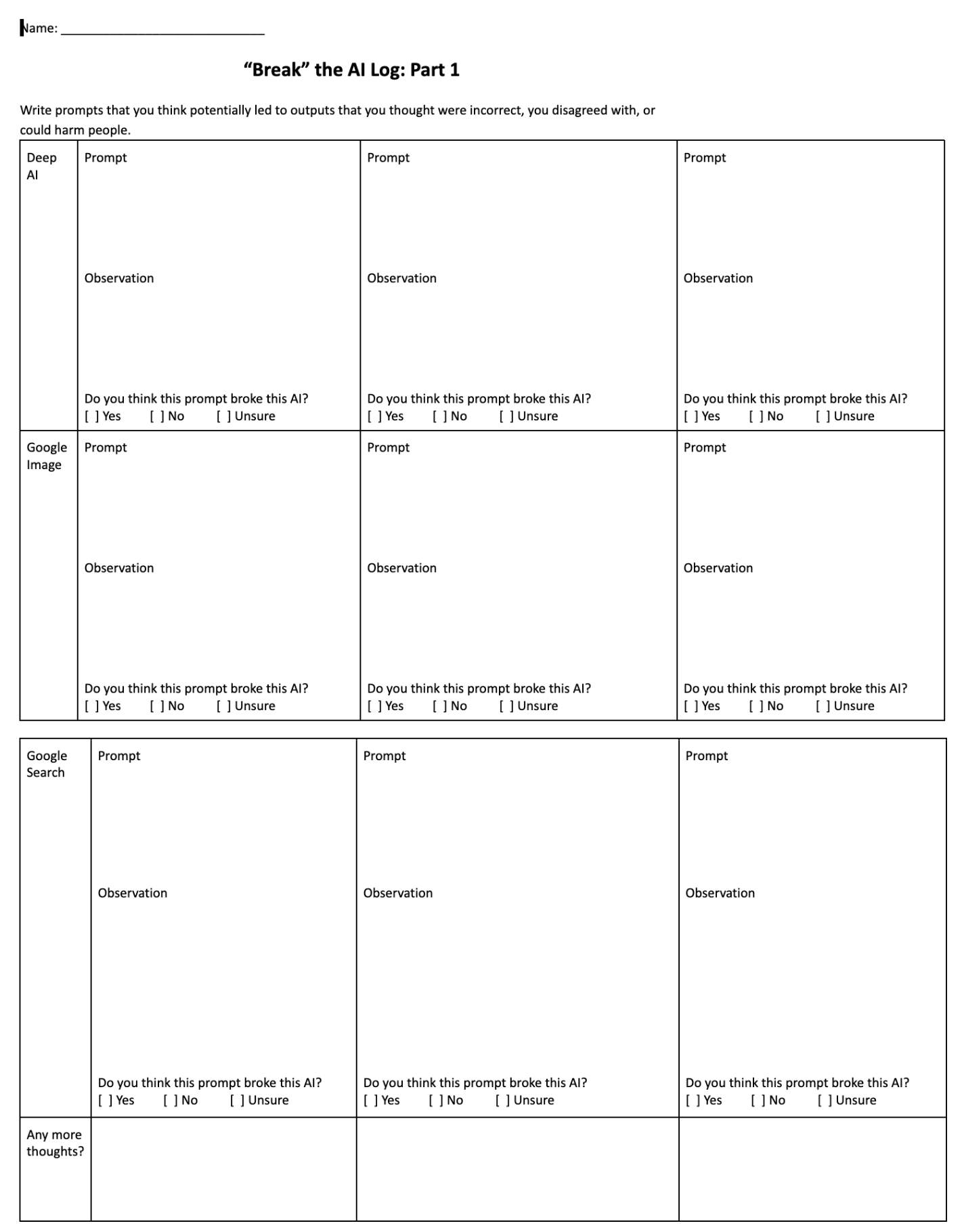}
    \caption{Break the AI Worksheet.}
    \label{fig:logpage1}
\end{figure}

\newpage
\subsection{Critical AI literacy intervention examples}
\label{appendix_criticalAILiteracy}
\begin{figure}[htbp]
    \centering
    \includegraphics[width=0.8\linewidth]{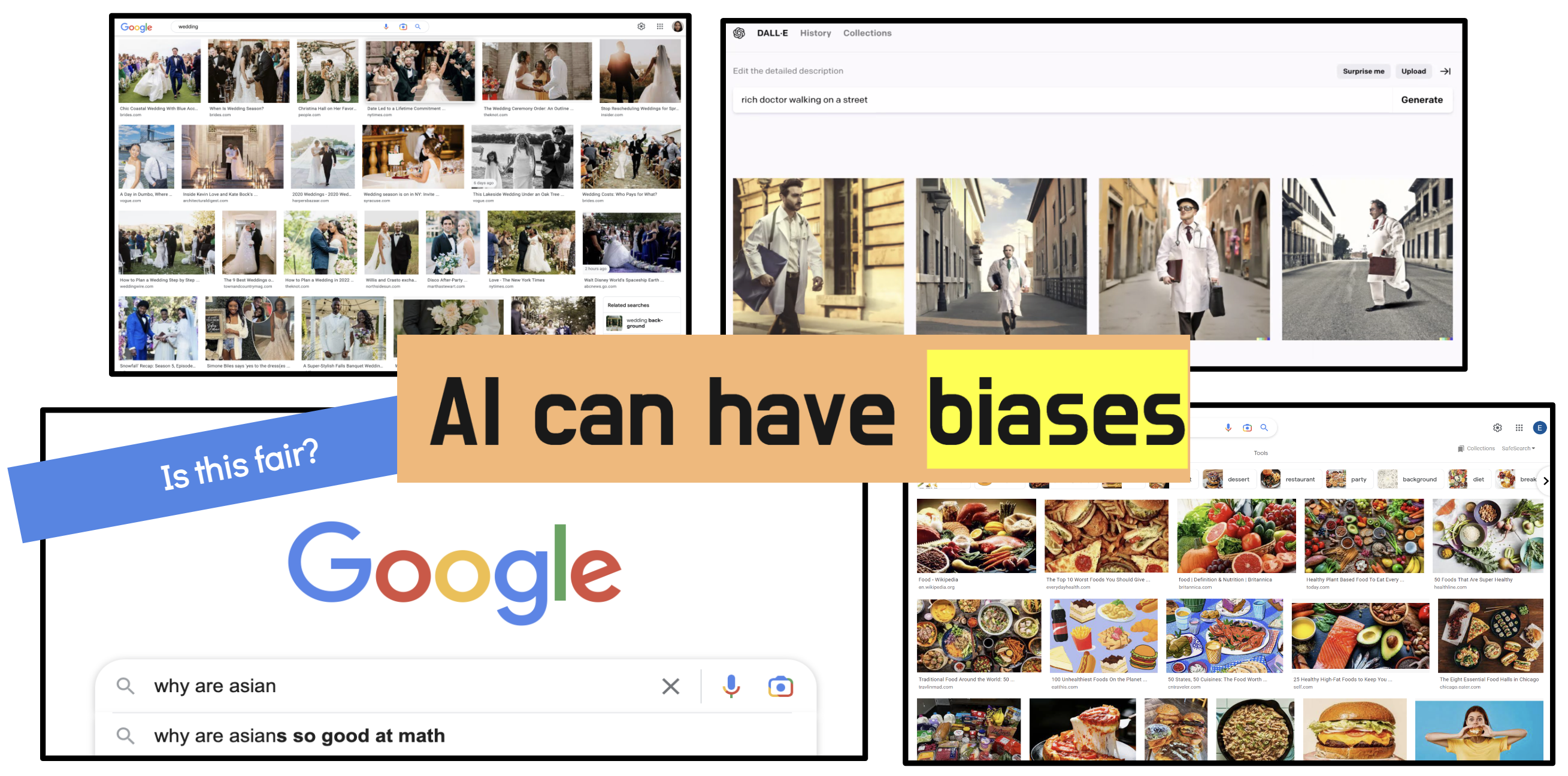}
    \caption{Slide demonstrating AI bias examples.}
    \label{fig:aibias}
\end{figure}

\subsection{Youth Auditing Tool interface}
\label{appendix_toolInterface}
\begin{figure}[htbp]
    \centering
    \includegraphics[width=0.8\linewidth]{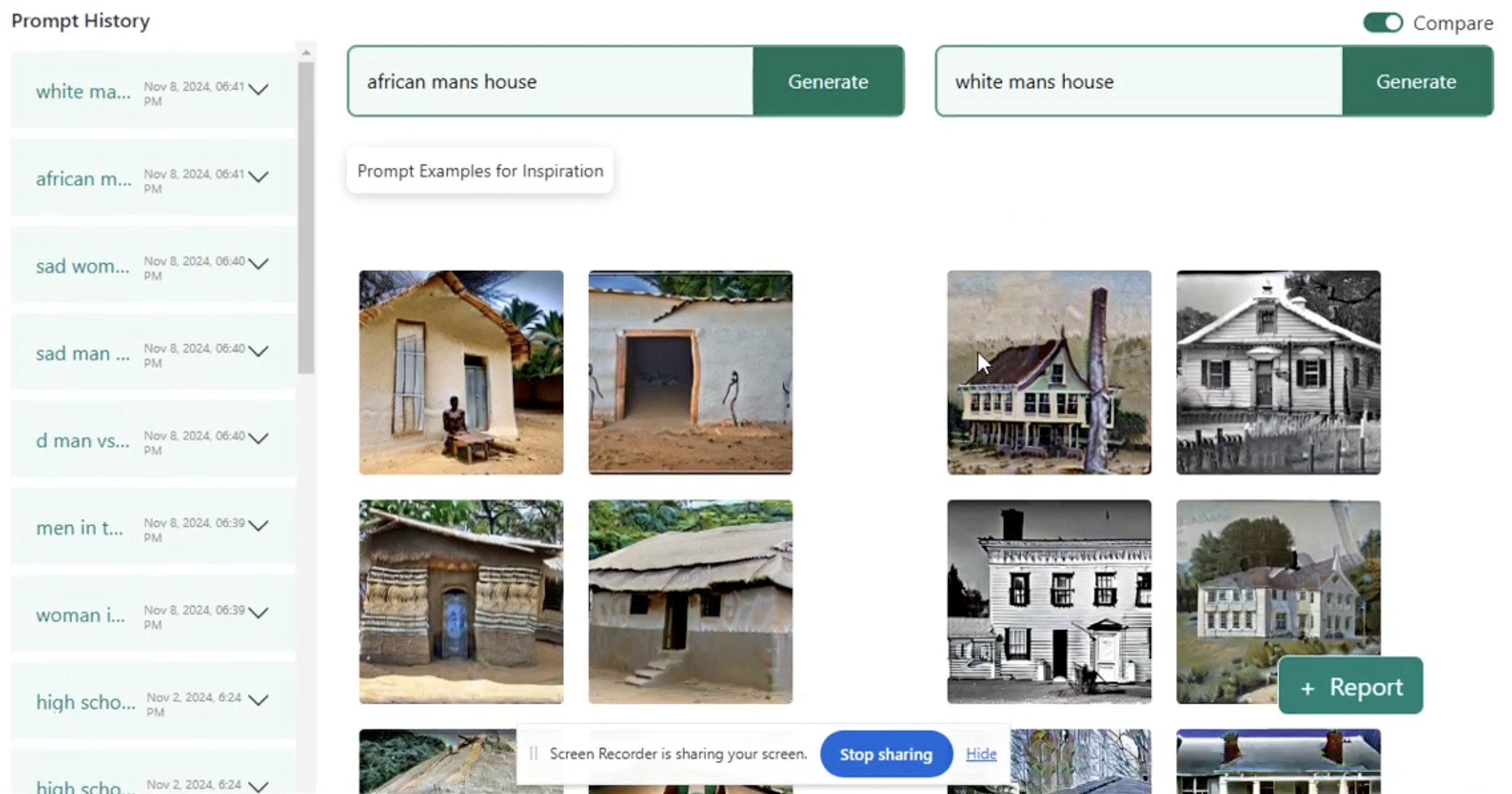}
    \caption{Audit tool interface.}
    \label{fig:tool}
\end{figure}


\newpage
\subsection{Example of Youth Auditing Tool onboarding slide}
\label{appendix_toolOnboarding}
\begin{figure} [h!]
    \centering
    \includegraphics[width=0.9\linewidth]{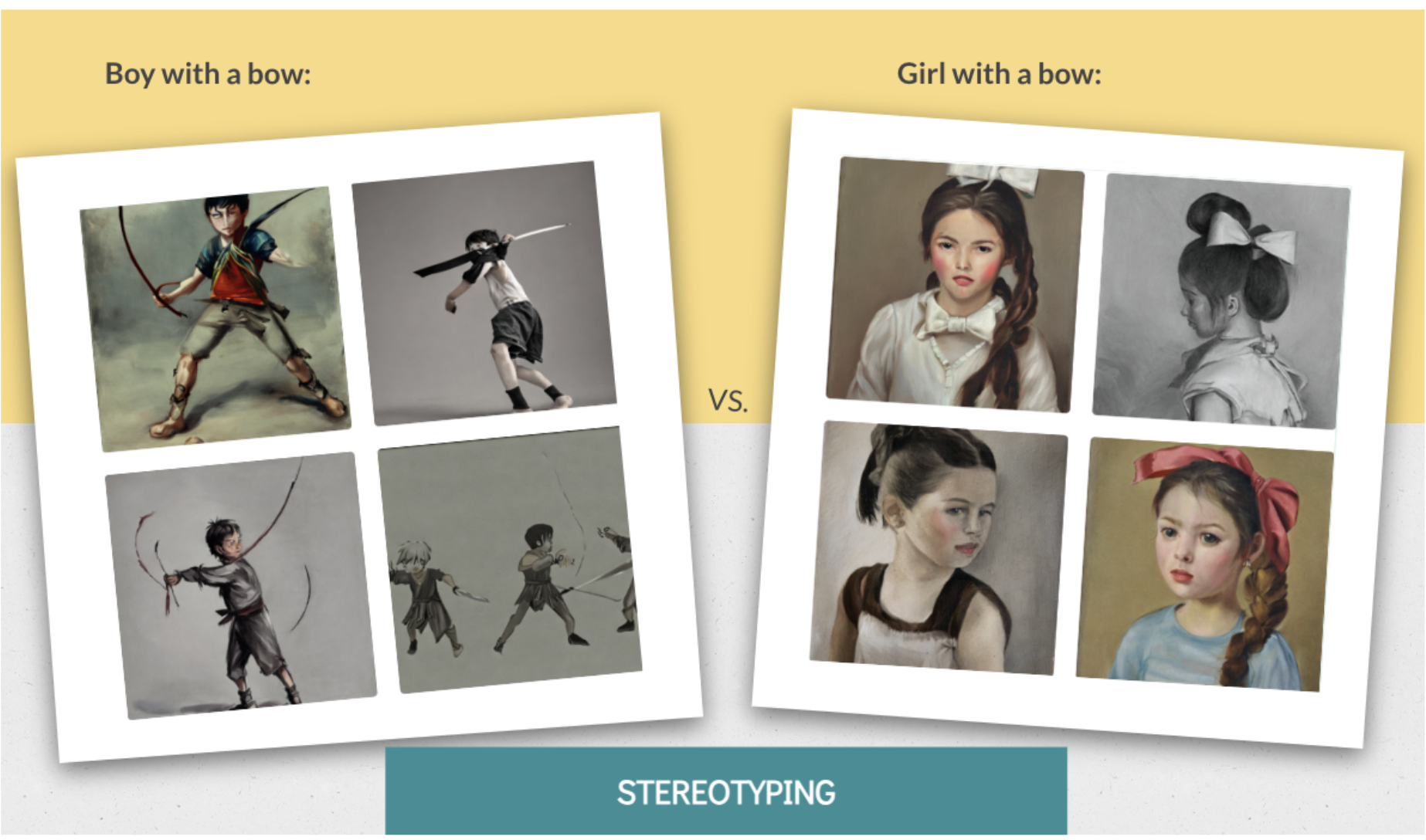}
    \caption{Audit tool results for boy with a bow versus girl with a bow.}
    \label{fig:toolresult2}
\end{figure}

\newpage
\subsection{Youth Auditing Tool report}
\label{appendix_toolReport}
\begin{figure}[htbp]
    \centering
    \includegraphics[width=0.8\linewidth]{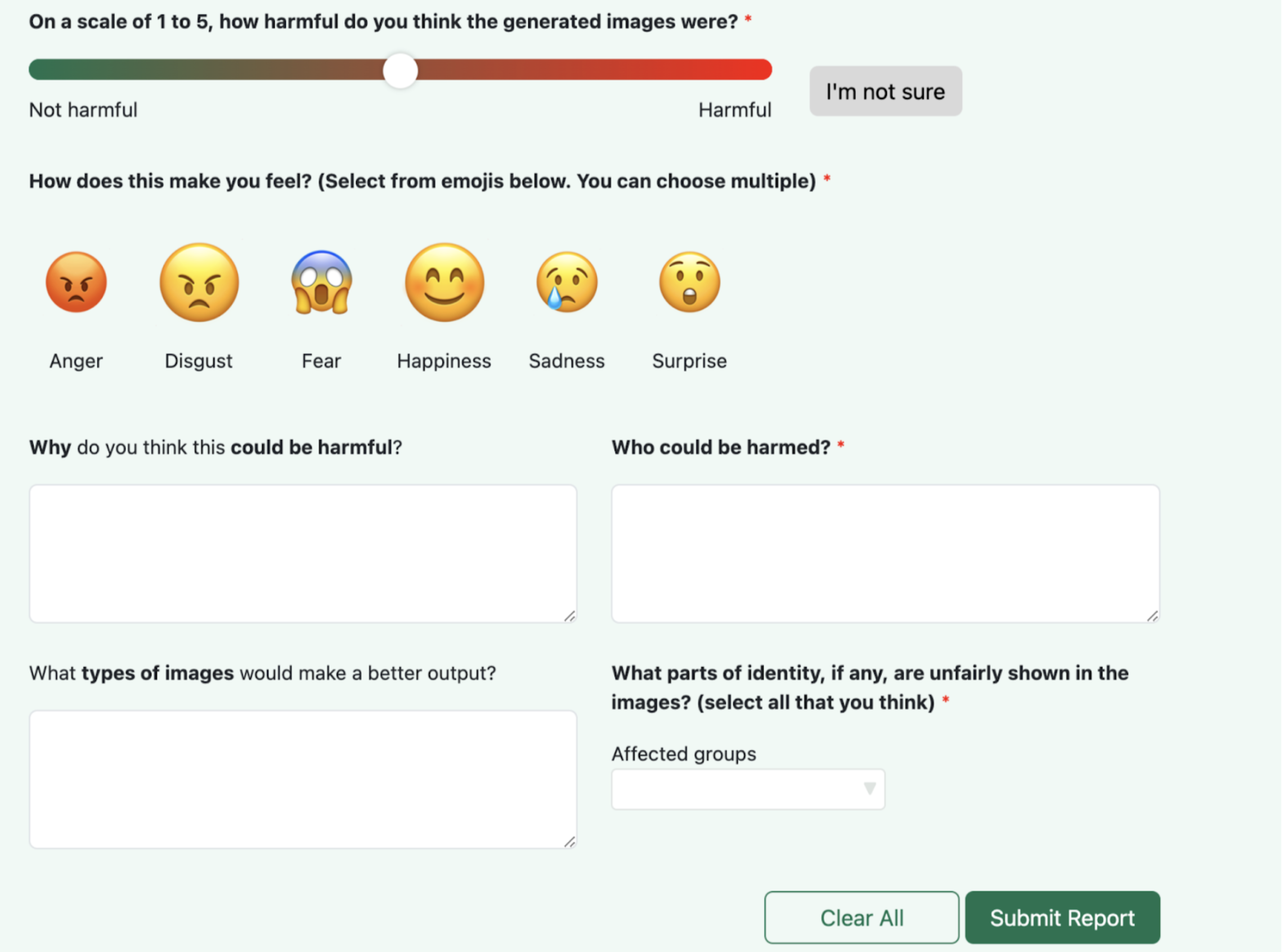}
    \caption{Audit tool report.}
    \label{fig:tool}
\end{figure}
\label{app:placeholder}

\end{document}